\documentstyle[preprint,aps]{revtex}

\begin{document}
\title{Comparative study of specific heat measurements in LaMnO$_{3}$, La$_{1.35}$Sr%
$_{1.65}$Mn$_{2}$O$_{7}$, La$_{1.5}$Sr$_{0.5}$NiO$_{4}$ and La$_{1.5}$Sr$%
_{0.5}$CoO$_{4}$}
\author{J. L\'{o}pez and O. F. de Lima}
\address{Instituto de F\'{i}sica Gleb Wataghin, Universidade Estadual de Campinas,\\
UNICAMP, 13083-970, Campinas, SP, Brazil}
\author{C. A. Cardoso and F. M. Araujo-Moreira}
\address{Depto. de F\'{i}sica, Universidade Federal de S\~{a}o Carlos, CP-676,\\
S\~{a}o Carlos, SP, 13565-905, Brazil}
\author{D. Prabhakaran}
\address{Department of Physics, Oxford University, Oxford OX1 3PU, United Kingdom}
\date{\today}
\maketitle

\begin{abstract}
We present the temperature dependence of the specific heat, without external
magnetic field and with {\em 9 T}, for LaMnO$_{3}$, La$_{1.35}$Sr$_{1.65}$Mn$%
_{2}$O$_{7}$, La$_{1.5}$Sr$_{0.5}$NiO$_{4}$ and La$_{1.5}$Sr$_{0.5}$CoO$_{4}$
single crystals. We found that spin-wave excitations in the ferromagnetic
and bilayer-structure La$_{1.35}$Sr$_{1.65}$Mn$_{2}$O$_{7}$ were suppressed
by the {\em 9 T} magnetic field. On the other hand, the external magnetic
field had no effect in the specific heat of the other three
antiferromagnetic samples. Also, the electronic part of the interactions
were removed at very low temperatures in the La$_{1.5}$Sr$_{0.5}$NiO$_{4}$
single crystal, even with a zero applied magnetic field. Below {\em 4 K}, we
found that the specific heat data for La$_{1.35}$Sr$_{1.65}$Mn$_{2}$O$_{7}$
and La$_{1.5}$Sr$_{0.5}$NiO$_{4}$ crystals could be fitted to an exponential
decay law. Detailed magnetization measurements in this low temperature
interval showed the existence of a peak close to 2 K. Both results,
magnetizations and specific heat suggest the existence of an anisotropy gap
in the energy spectrum of La$_{1.35}$Sr$_{1.65}$Mn$_{2}$O$_{7}$ and La$%
_{1.5} $Sr$_{0.5}$NiO$_{4}$ compounds.
\end{abstract}

\section{\protect\bigskip Introduction}

The combined ordering of charge (CO) and spin (SO) is proving to be a common
phenomenon in transition-metal oxides like LaMnO$_{3}$, La$_{1.35}$Sr$%
_{1.65} $Mn$_{2}$O$_{7}$, La$_{1.5}$Sr$_{0.5}$NiO$_{4}$ and La$_{1.5}$Sr$%
_{0.5}$CoO$_{4}$. The magnetic properties of Mn, Co and Ni perovskites are
considered to arise from the strong competition involving ferromagnetic and
antiferromagnetic interactions and the spin-phonon coupling\cite{Radaelli}$-$%
\cite{J.López2}. The dimensionality of the relevant structure involving the
transition metal ions, three (3D) or quasi-two-dimensional (2D), also plays
an important role. For instance, the bilayer-structure compounds La$_{2-2x}$%
Sr$_{1+2x}$Mn$_{2}$O$_{7}$, in which MnO$_{2}$ and (La,Sr)$_{2}$O$_{2}$
layers are stacked alternatively have 2D electronic and magnetic properties 
\cite{Okuda}.

The well known LaMnO$_{3}$ is an antiferromagnetic insulator with 3D
characteristics. The specific heat at low temperature, for LaMnO$_{3+\delta
} $ samples, was found by L. Ghivelder et al.\cite{Ghivelder} to be very
sensitive to small variations of $\delta $. Previous specific heat
measurements in the related Nd$_{0.5}$Sr$_{0.5}$MnO$_{3}$, Nd$_{0.5}$Ca$%
_{0.5}$MnO$_{3}$, Sm$_{0.5}$Ca$_{0.5}$MnO$_{3}$, Dy$_{0.5}$Ca$_{0.5}$MnO$%
_{3} $ and Ho$_{0.5}$Ca$_{0.5}$MnO$_{3}$ samples revealed a Schottky-like
anomaly at low temperatures\cite{JLópez3}$^{,}$\cite{JLópez4}. Differently
from the LaMnO$_{3}$ crystal, the La$_{1.5}$Sr$_{0.5}$NiO$_{4}$ and La$%
_{1.5} $Sr$_{0.5}$CoO$_{4}$ compounds have their magnetic ions (Ni and Co)
confined in planes which are insulated by (La,Sr)$_{2}$O$_{2}$ layers. I. A.
Zaliznyak et al.\cite{Zaliznyak} presented elastic and quasielastic neutron
scattering measurements characterizing peculiar short-range charge-orbital
and spin order in the layered perovskite La$_{1.5}$Sr$_{0.5}$CoO$_{4}$. They
found that, below {\em T}$_{c}${\em \ = 750 K}, holes introduced by Sr
doping lose mobility and enter into a statically ordered charge-glass-phase
with loosely correlated checkerboard arrangement of empty and occupied d$%
_{3z^{2}-r^{2}}$ orbitals (Co$^{3+}$ and Co$^{2+}$). La$_{1.5}$Sr$_{0.5}$NiO$%
_{4}$, like its parent compound La$_{2}$NiO$_{4}$, is an insulator, contrary
to the related compound La$_{2}$CuO$_{4}$, where the antiferromagnetic
insulator phase is rapidly destroyed by doping, leading to a metallic
superconductor phase at moderate hole concentration\cite{Sachan}, La$_{2}$NiO%
$_{4}$ remains nonmetallic up to quite large hole concentrations\cite
{Wochner}. R. Kajimoto et al. \cite{Kajimoto} studied the CO in La$_{1.5}$Sr$%
_{0.5}$NiO$_{4}$ with neutron diffraction technique. They found a
rearrangement of CO from checkerboard-type to stripe-type as a function of
temperature. In their measurements the stripe phase persisted up to x = 0.7
for highly hole-doped samples of Nd$_{2-x}$Sr$_{x}$NiO$_{4}$ with 0.45 $\leq 
$ x $\leq $ 0.7.

A large number of papers discuss the properties of Mn, Co and Ni perovskite
compounds treated separately. However, to our knowledge, a comparative study
of the low temperature specific heat for these perovskite families is
missing. Here, we present the temperature dependence of the specific heat,
without external magnetic field ({\em H}) and with {\em H=9 T}, for LaMnO$%
_{3}$, La$_{1.35}$Sr$_{1.65}$Mn$_{2}$O$_{7}$, La$_{1.5}$Sr$_{0.5}$NiO$_{4}$
and La$_{1.5}$Sr$_{0.5}$CoO$_{4}$ single crystals. We found that the
specific heat data for La$_{1.35}$Sr$_{1.65}$Mn$_{2}$O$_{7}$ and La$_{1.5}$Sr%
$_{0.5}$NiO$_{4}$ crystals could be fitted to an exponential decay law below 
{\em 4 K}. Detailed magnetization measurements in this low temperature
interval showed the existence of a peak close to 2 K. Both results,
magnetization and specific heat suggest the existence of an anisotropy gap
in the energy spectrum of La$_{1.35}$Sr$_{1.65}$Mn$_{2}$O$_{7}$ and La$%
_{1.5} $Sr$_{0.5}$NiO$_{4}$ compounds. Our macroscopical measurements
confirm the complex magnetic excitation and electronic band structure due to
charge ordering and the quasi-2D confinement of the magnetic ions.

\section{Experimental methods}

Large single crystals of LaMnO$_{3}$, La$_{1.35}$Sr$_{1.65}$Mn$_{2}$O$_{7}$,
La$_{1.5}$Sr$_{0.5}$NiO$_{4}$ and La$_{1.5}$Sr$_{0.5}$CoO$_{4}$ were grown
by the floating zone method described elsewhere\cite{Prabhakaran}. The
magnetization measurements were done with a Quantum Design MPMS-5S SQUID
magnetometer. Specific heat measurements were made with a Quantum Design
PPMS calorimeter that uses a {\it two relaxation times} technique, and data
was always collected during sample cooling. The intensity of the heat pulse
applied to the sample was calculated to produce a variation in the
temperature bath of {\em 0.5 \%}. Experimental errors during the specific
heat and magnetization measurements were lower than {\em 1 \%} for all
temperatures and samples.

\section{Results and Discussion}

Specific heat measurements at low temperatures give valuable information
about the ground state excitations. In contrast to magnetization, which has
a vector character, the specific heat is an scalar property. Figures 1a, 1b,
1c and 1d show the dependence of the specific heat measurements with
temperature, between 2 and 30 K, for LaMnO$_{3}$, La$_{1.35}$Sr$_{1.65}$Mn$%
_{2}$O$_{7}$, La$_{1.5}$Sr$_{0.5}$NiO$_{4}$ and La$_{1.5}$Sr$_{0.5}$CoO$_{4}$
single crystals, studied with {\em H=0} and {\em 9 T}. The data is plotted
as {\em C/T} vs. {\em T}$^{2}$ to facilitate the interpretation. The LaMnO$%
_{3}$ and La$_{1.5}$Sr$_{0.5}$CoO$_{4}$ samples showed both an almost linear
behavior and no magnetic field dependence. On the other hand, the magnetic
field strongly affected the low temperature ({\em T 
\mbox{$<$}%
5 K}) behavior of the La$_{1.35}$Sr$_{1.65}$Mn$_{2}$O$_{7}$ compound. The
specific heat curves for the La$_{1.5}$Sr$_{0.5}$NiO$_{4}$ sample did not
change with the applied field, but were not linear at temperatures below 
{\em 5 K}.\bigskip\ Continuous lines in figure 1 indicate the fitting of the
experimental data between 2 and 30 K by the following expression\cite{Gordon}%
:

\begin{equation}
C=\sum \beta _{2n+1}\,T^{\text{\/}2n+1}{\em +\ }\beta _{3/2}\,T^{\text{\/}%
3/2}  \label{5}
\end{equation}
The whole temperature interval, from {\em 2 K} to {\em 30 K}, was possible
to be fitted with natural values of $n$ from {\em 0} to {\em 4}. ${\em C}$
is the specific heat, ${\em T}$ is the temperature and $\beta $ parameters
represent the contributions of electron interactions ($\beta _{1}$),
ferromagnetic spin waves ($\beta _{3/2}$) and phonon modes ($\beta _{3}$, $%
\beta _{5}$, $\beta _{7}$ and $\beta _{9}$). The coefficient $\beta _{1}$ (%
{\it n=0}) is also known as $\gamma $, and $\beta _{3}$\ ({\it n=1}) as $%
\beta $. The results of the specific heat data fitting are shown in Table 1.

It is worth stressing that, differently from the LaMnO$_{3}$ crystal, the La$%
_{1.5}$Sr$_{0.5}$NiO$_{4}$ and La$_{1.5}$Sr$_{0.5}$CoO$_{4}$ compounds have
their magnetic ions (Ni and Co) confined in planes which are insulated by
(La,Sr)$_{2}$O$_{2}$ layers. However, these latter two compounds show a
different magnetic behavior compared to the quasi-2D La$_{1.35}$Sr$_{1.65}$Mn%
$_{2}$O$_{7}$ sample: the\ Mn\ ions order ferromagnetically, while the
corresponding Ni and Co ions order antiferromagnetically. Therefore, as seen
in figures 1c and 1d, there is not much contribution of the magnetic field
into the measured specific heat.

>From resistivity measurements the LaMnO$_{3}$, La$_{1.5}$Sr$_{0.5}$NiO$_{4}$
and La$_{1.5}$Sr$_{0.5}$CoO$_{4}$ samples are electrical insulators at low
temperatures, and applied magnetic fields up to 9 T are not strong enough to
destroy this characteristic\cite{Zaliznyak},\cite{Wochner},\cite{Myron}.
Therefore, we should not expect for the previous three crystals the linear
contribution from free electrons to the specific heat. However, other kind
of many-body excitations could also lead to a linear contribution\cite
{Smolyaninova1}. On the other hand, the La$_{1.35}$Sr$_{1.65}$Mn$_{2}$O$_{7}$
is metallic below about 100 K\cite{Okuda}. Our fitting shows that $\beta
_{1} $ values are big for La$_{1.35}$Sr$_{1.65}$Mn$_{2}$O$_{7}$ and La$_{1.5}
$Sr$_{0.5}$NiO$_{4}$. To facilitate even more the comparison, we have
re-plotted in figure 2 the data for the La$_{1.35}$Sr$_{1.65}$Mn$_{2}$O$_{7}$
and La$_{1.5}$Sr$_{0.5}$NiO$_{4}$ crystals at temperatures below 10 K.
Closed symbols represent the measurements with {\em H=0} and open ones with 
{\em 9 T}.

Okuda et al.\cite{Okuda} found that the decrease of specific heat at low
temperatures, due to an applied magnetic field of 9 T, was about ten times
larger in a La$_{1.35}$Sr$_{1.65}$Mn$_{2}$O$_{7}$ sample than the observed
values in La$_{1-x}$Sr$_{x}$MnO$_{3}$ samples (with x=0.3 and 0.4). They
also calculated the theoretical reduction in specific heat upon application
of a magnetic field for the ideal simple-cubic (3D) and simple-square (2D)
lattices and concluded that the observed change in specific heat for the
bilayered manganite was large, but still less than that for an ideal 2D
ferromagnetism. Besides, Okuda et al.\cite{Okuda} reported values of $\beta
_{1}$=3 mJ/mole-K$^{2}$ for a La$_{1.35}$Sr$_{1.65}$Mn$_{2}$O$_{7}$\ sample,
an order of magnitude smaller than the one found by us. Because their $\beta
_{1}$ value was similar to the one found in three dimensional perovskites,
like La$_{0.7}$Sr$_{0.3}$MnO$_{3}$\cite{Okuda0}, they concluded that
dimensionality did not affect the value of the electron-electron interaction
constant. However, our result shows that the quasi-2D confinement of the
electrons in La$_{1.35}$Sr$_{1.65}$Mn$_{2}$O$_{7}$ do increase the
electron-electron interaction constant in comparison to the 3D counterpart.

As expected, a term of the type T$^{3/2}$ appears only for the La$_{1.35}$Sr$%
_{1.65}$Mn$_{2}$O$_{7}$\ sample due to its ferromagnetic interactions. B. F.
Woodfield et al.\cite{Woodfield} studied the specific heat in La$_{1-x}$Sr$%
_{x}$MnO$_{3}$ with {\em x} between 0.1 and 0.3 and found $\beta _{3/2}$
values in the interval 0.9 to 3.7 mJ/mole-K$^{5/2}$. Our $\beta _{3/2}$
values, for La$_{1.35}$Sr$_{1.65}$Mn$_{2}$O$_{7}$, are approximately equals
to the 3D counterpart and roughly duplicate upon the application of a 9 T
magnetic field. Figure 1c (also in fig 2) shows that the La$_{1.5}$Sr$_{0.5}$%
NiO$_{4}$ curves make a downward turn at low temperatures in both {\em H=0}
and {\em 9 T} and this region is not well fitted by equation 1. The values
of $\beta _{1}$ for this sample (see table 1) apply only for temperatures
higher than 4 K, and reveal a very high electronic interactions.

Martinho et al. \cite{Martinho} interpreted the specific heat of La$_{2-2x}$%
Sr$_{1+2x}$Mn$_{2}$O$_{7}$\ samples ({\em 0.29%
\mbox{$<$}%
x 
\mbox{$<$}%
0.51}) as thermal excitations of a two-dimensional gas of ferromagnetic
magnons. However, they did not discuss the very low temperature interval
(below 4 K). In figure 3 we re-scale the specific heat data in the
temperature interval below {\em 4 K} for the La$_{1.35}$Sr$_{1.65}$Mn$_{2}$O$%
_{7}$\ and La$_{1.5}$Sr$_{0.5}$NiO$_{4}$ crystals. The x-axis is now equal
to the inverse of temperature and the y-axis in presented in logarithmic
scale to facilitate the comparison with an exponential decay law. In both, 
{\em H=0} and {\em 9 T}, the data points are well fitted by straight lines.
The estimated energy gap ({\em E}$_{gap}$) in the La$_{2-2x}$Sr$_{1+2x}$Mn$%
_{2}$O$_{7}$\ crystal was {\em 0.30 meV} and {\em 0.57 meV} for zero and 9
T, respectively. On the other hand, the estimated {\em E}$_{gap}$ in the La$%
_{1.5}$Sr$_{0.5}$NiO$_{4}$ crystal was {\em 0.63 meV} and {\em 0.65 meV} for
zero and {\em 9 T}, respectively. The graph for La$_{1.5}$Sr$_{0.5}$NiO$_{4}$
remind us the structure-related superconductor La$_{1.5}$Sr$_{0.5}$CuO$_{4}$%
. In a superconductor an exponential decay in the specific heat is
interpreted as the opening of a gap in the electronic structure\cite{Kittel}%
. However, our sample, differently from superconductors, did not show a
noticeable dependence with a magnetic field up to {\em 9 T}. If a BCS-like
theory\cite{Kittel} were to be valid in this crystal ({\em E}$_{gap}${\em \
= 7/2 k}$_{B}${\em \ T}$_{c}$) the sample should have a corresponding
critical temperature ({\em T}$_{c}$) at about {\em 2 K}.

Figure 4\ shows a detailed measurement of the zero field cooling ({\em ZFC})
magnetization in the La$_{1.35}$Sr$_{1.65}$Mn$_{2}$O$_{7}$ crystal, between
1.8 and 10 K, with applied magnetic field of 20 Oe (a) and 50 Oe (b).
Measurements were done with the magnetic field parallel (triangles) and
perpendicular (squares) to the {\em c} axes and temperature steps of {\em %
0.1 K}. The results reveal a clear anisotropy due to the quasi-bidimensional
distribution of the magnetic ions. The magnetization shows three features: a
minimum close to 2 K, a small maximum close to 4 K and a plateau close to 7
K. Figure 5\ shows {\em ZFC }magnetization measurement in the La$_{1.5}$Sr$%
_{0.5}$NiO$_{4}$ crystal between 1.8 and 10 K with applied magnetic field of
1 T (a) and 3 T (b). The magnetic field was applied parallel (triangles) and
perpendicular (squares) to the {\em c} axes and the temperature steps were
of {\em 0.1 K}. These graphs also display an anisotropy behavior due to the
quasi-bidimensional distribution of the magnetic ions. The magnetization
here shows two features: a maximum close to 2 K and a minimum close to 4 K
in the orientation of the applied magnetic field parallel to the {\em c}
axes. The absolute value of magnetization is higher in La$_{1.35}$Sr$_{1.65}$%
Mn$_{2}$O$_{7}$, due to its ferromagnetic alignment, in comparison with the
antiferromagnetic alignment in La$_{1.5}$Sr$_{0.5}$NiO$_{4}$.\ In both cases
the position of the peaks close to 2 K in the magnetization curves (figures
4 and 5) seems to be correlated to the exponential decay in the specific
heat (figure 3).

Recently, Boothroyd\cite{Boothroyd} et al. studied a single crystal of La$%
_{1.5}$Sr$_{0.5}$NiO$_{4}$\ with polarized neutrons at 10 K. They made
neutron energy scans with the direction of the incident beam fixed and
aligned parallel, as well as perpendicular, to the NiO layers. They found
inter-plane correlations at low energies (negligible for E $\geq $ 5 meV)
and a reduction in intensity below an energy of 4 meV. Given that neutrons
scatter from spin fluctuations perpendicular to the wave vector, these
observations indicated that the intensity reduction below 4 meV was due to
the freezing out of the c-axis component of the spin fluctuations. They also
pointed out the existence of a 4 meV energy gap due to single-ion
out-of-plane anisotropy. Although we found in the La$_{1.5}$Sr$_{0.5}$NiO$%
_{4}$ crystal, using specific heat measurements, a gap value smaller than
the one reported by Boothroyd et al.\cite{Boothroyd}, both results, one
microscopically and the other macroscopically, seem to confirm the complex
magnetic excitation and band structure due to charge ordering and the
quasi-2D confinement of the Ni ions. Further studies are clearly necessary
to elucidate better these points.

\section{Conclusions}

Single crystals of LaMnO$_{3}$, La$_{1.35}$Sr$_{1.65}$Mn$_{2}$O$_{7}$, La$%
_{1.5}$Sr$_{0.5}$NiO$_{4}$ and La$_{1.5}$Sr$_{0.5}$CoO$_{4}$ were
characterized by magnetization and specific heat measurements. The bilayer
compound La$_{1.35}$Sr$_{1.65}$Mn$_{2}$O$_{7}$ presents a ferromagnetic
transition, while the other studied compositions show an antiferromagnetic
order. Spin-wave excitations in the bilayer-structure La$_{1.35}$Sr$_{1.65}$%
Mn$_{2}$O$_{7}$ are suppressed by a 9 T magnetic field as indicated by
specific heat measurements. This is attributed to the reduced magnetic
dimensionality. The effect is larger than in the case of 3D compounds, but
not as large as expected in an ideal 2D system. Unlike a previous report\cite
{Okuda}, we found that the quasi-2D confinement of the electrons in La$%
_{1.35}$Sr$_{1.65}$Mn$_{2}$O$_{7}$ do increase the electron-electron
interaction constant in comparison to its 3D counterpart. The specific heat
does not change with a 9 T magnetic field in LaMnO$_{3}$, La$_{1.5}$Sr$%
_{0.5} $NiO$_{4}$ and La$_{1.5}$Sr$_{0.5}$CoO$_{4}$. However, electronic
excitations are drastically removed at very low temperatures in the La$%
_{1.5} $Sr$_{0.5}$NiO$_{4}$ single crystal, as revealed by the downward turn
in the specific heat. Below {\em 4 K}, we also found that the specific heat
data for the La$_{1.35}$Sr$_{1.65}$Mn$_{2}$O$_{7}$ and La$_{1.5}$Sr$_{0.5}$%
NiO$_{4}$ crystals could be fitted by an exponential decay law. From the
fittings we were able to estimate characteristic energy gaps for both
compounds. Detailed magnetization measurements in this low temperature
interval showed the existence, close to 2 K, of a maximum for La$_{1.5}$Sr$%
_{0.5}$NiO$_{4}$ and a minimum for La$_{1.35}$Sr$_{1.65}$Mn$_{2}$O$_{7}$.
Our measurements of magnetizations and specific heat, combined with a
previous report on neutron diffraction\cite{Boothroyd}, suggest the
existence of an anisotropy gap in the energy spectrum of the La$_{1.35}$Sr$%
_{1.65}$Mn$_{2}$O$_{7}$ and La$_{1.5}$Sr$_{0.5}$NiO$_{4}$ compounds.

\section{Acknowledgments}

We thank the Brazilian science agencies FAPESP (Funda\c{c}\~{a}o de Amparo
\`{a} Pesquisa do Estado de S\~{a}o Paulo) and CNPq (Conselho Nacional de
Desenvolvimento Cient\'{i}fico e Tecnol\'{o}gico) for financial support.

\bigskip

\bigskip \newpage

Table 1. Results of the fitting to the law $C=\sum \beta _{2n+1}\,T^{\text{\/%
}2n+1}${\em \ + }$\beta _{3/2}\,T^{\text{\/}3/2}$, with {\em n} from 0 to 4,
for the four studied single crystals and {\em H= 0} and {\em 9 T}. The units
are mJ/mole-K$^{j}$, where {\em j} is the subscript of the coefficient.

\bigskip

\bigskip 
\begin{tabular}{|l|l|l|l|l|l|l|l|}
\hline
Sample & H ( T ) & $\beta _{1}$ & $\beta _{3/2}$ & $\beta _{3}$ & $\beta
_{5} $ & $\beta _{7}$ & $\beta _{9}$ \\ \hline
LaMnO$_{3}$ & 0 & 1.1 & 0 & 0.20 & 0.31 & -0.48 & 2.0 \\ \hline
& 9 & 1.1 & 0 & 0.20 & 0.31 & -0.48 & 2.0 \\ \hline
La$_{1.35}$Sr$_{1.65}$Mn$_{2}$O$_{7}$ & 0 & 24 & 2.4 & 0.12 & 1.3 & -1.4 & 
5.3 \\ \hline
& 9 & 14 & 4.4 & 0.20 & 0.98 & -1.0 & 3.5 \\ \hline
La$_{1.5}$Sr$_{0.5}$NiO$_{4}$ & 0 & 16 & 0 & 0.18 & 0.67 & -0.79 & 2.8 \\ 
\hline
& 9 & 16 & 0 & 0.17 & 0.71 & -0.82 & 2.9 \\ \hline
La$_{1.5}$Sr$_{0.5}$CoO$_{4}$ & 0 & 1.4 & 0 & 0.12 & 0.61 & -0.76 & 2.9 \\ 
\hline
& 9 & 1.9 & 0 & 0.12 & 0.61 & -0.76 & 2.9 \\ \hline
\end{tabular}
\bigskip

\bigskip \newpage

Figure Captions

\bigskip

\bigskip

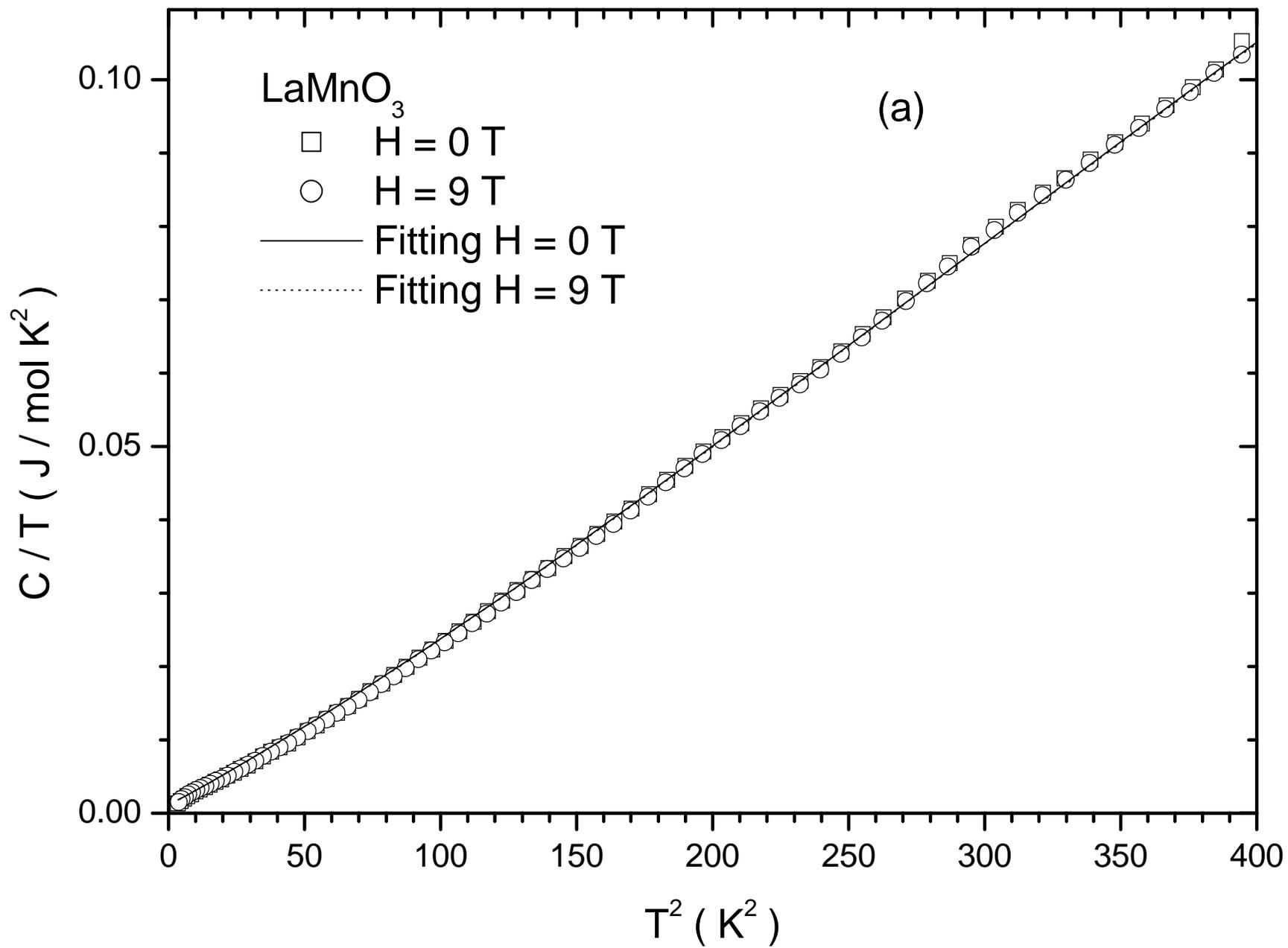
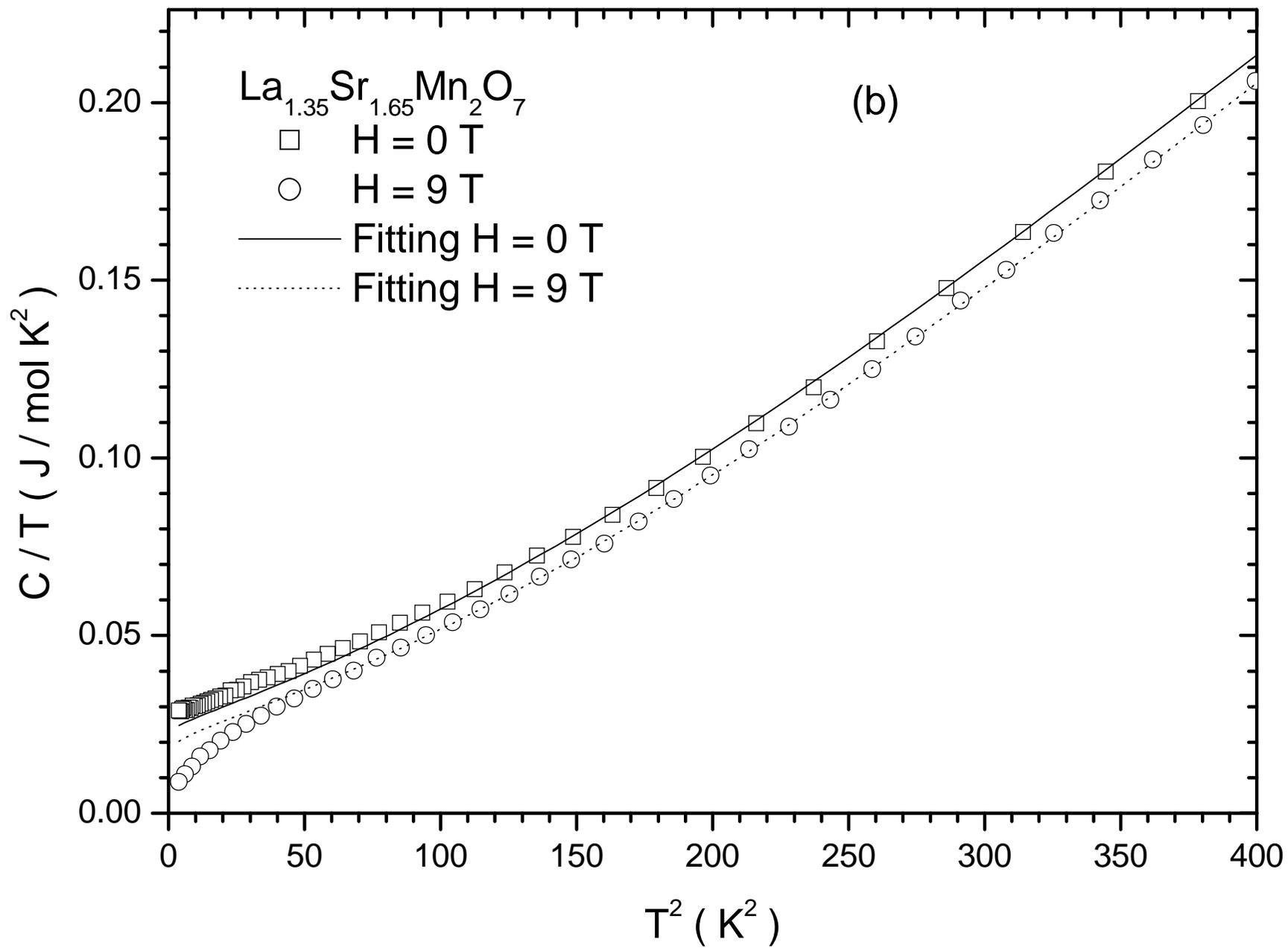
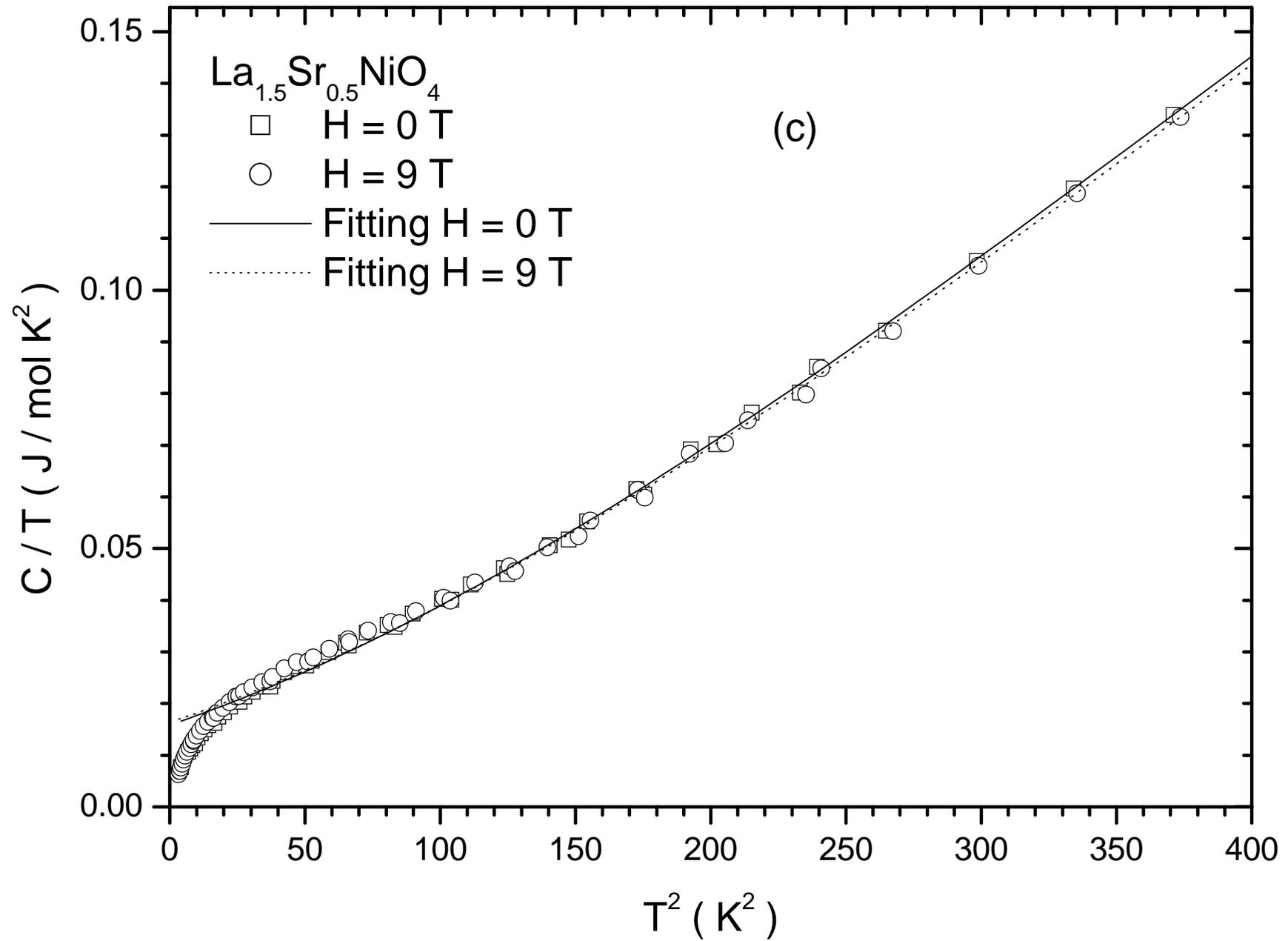
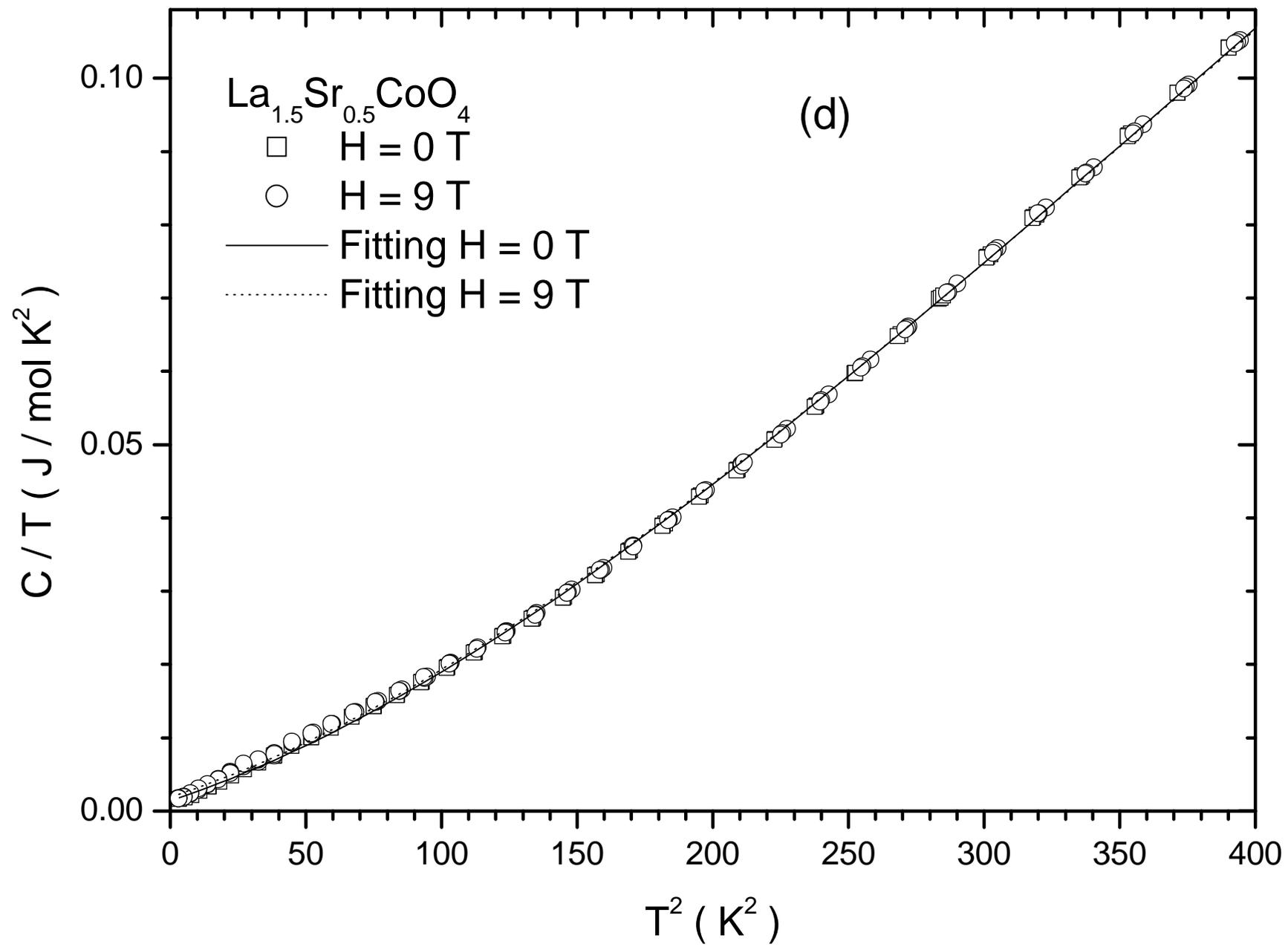
Figure 1. Specific heat measurements between 2 and 30 K in the single
crystals LaMnO$_{3}$, La$_{1.35}$Sr$_{1.65}$Mn$_{2}$O$_{7}$, La$_{1.5}$Sr$%
_{0.5}$NiO$_{4}$ and La$_{1.5}$Sr$_{0.5}$CoO$_{4}$, with {\em H=0} and {\em %
9 T}. Continuous lines represent the fitting of the law: $C=\sum \beta
_{2n+1}\,T^{\text{\/}2n+1}${\em \ + }$\beta _{3/2}\,T^{\text{\/}3/2}$. The
data is plotted as {\em C/T} vs. {\em T}$^{2}$ to facilitate the
interpretation.

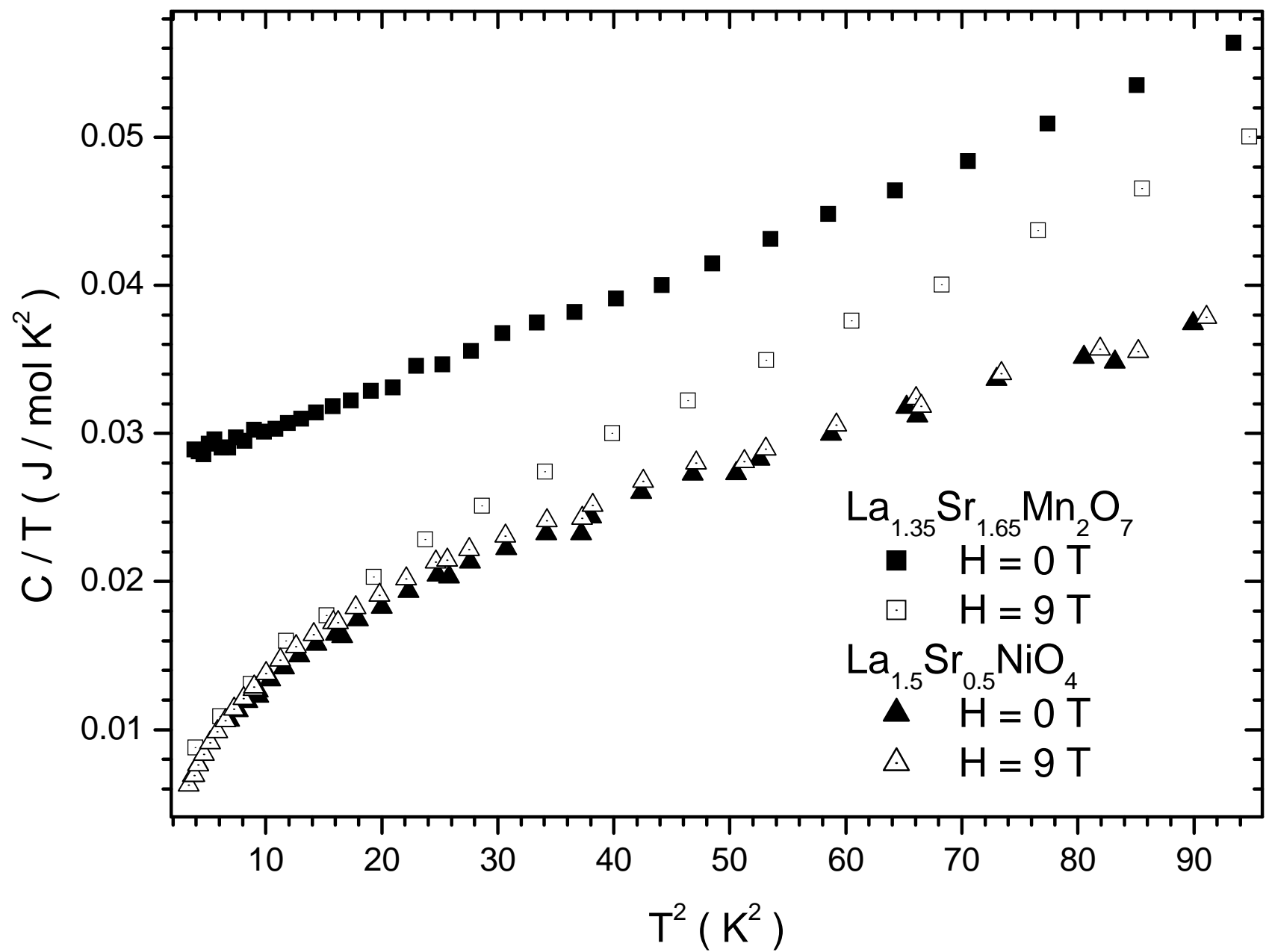
Figure 2. Specific heat of La$_{1.35}$Sr$_{1.65}$Mn$_{2}$O$_{7}$ and La$%
_{1.5}$Sr$_{0.5}$NiO$_{4}$ crystals below 10 K. Close symbols represent the
measurements with {\em H=0} and open ones with {\em 9 T}.

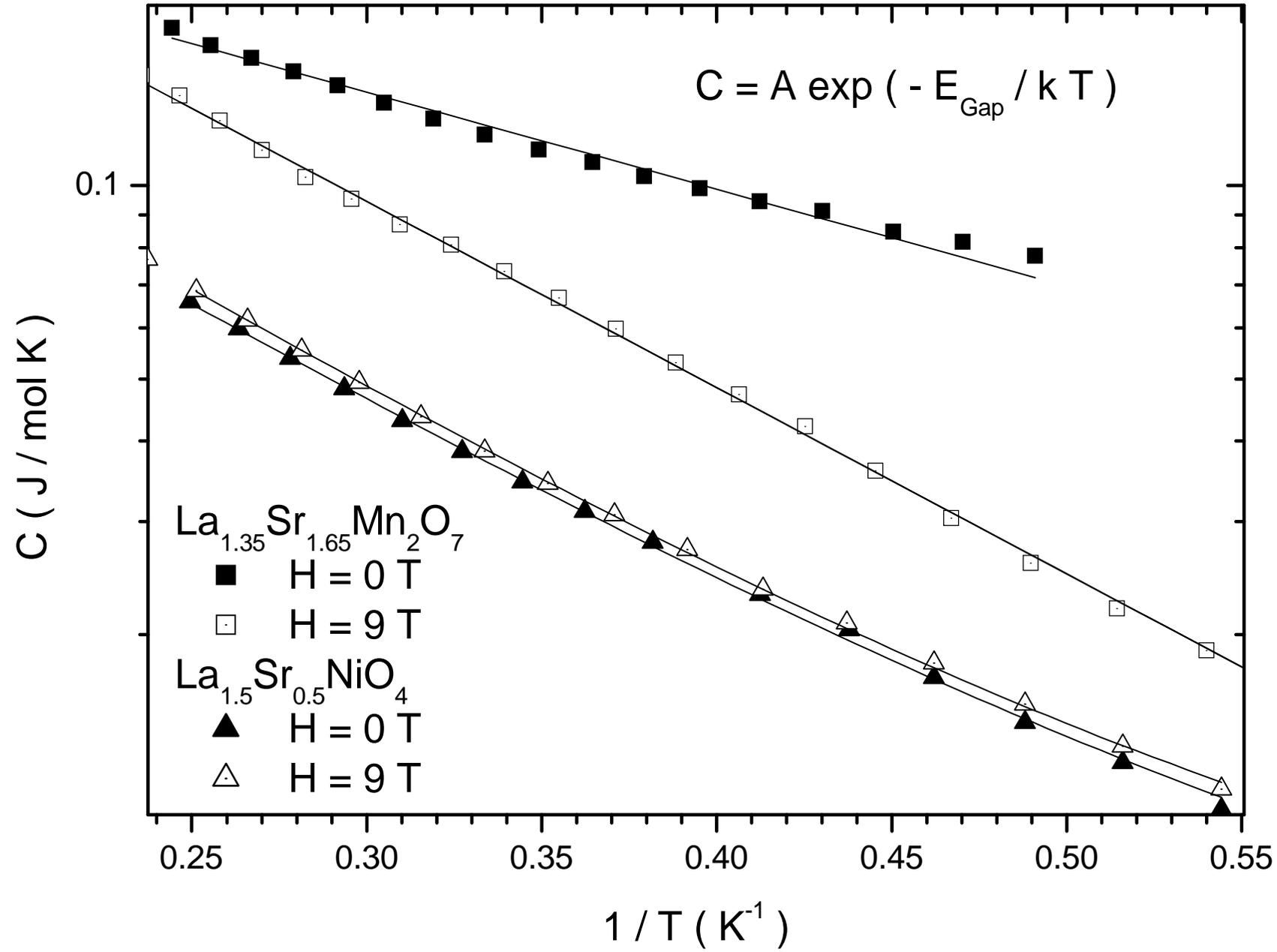
Figure 3. Re-scaled specific heat data in the low temperature interval
(below {\em 4 K}) for the La$_{1.35}$Sr$_{1.65}$Mn$_{2}$O$_{7}$\ and La$%
_{1.5}$Sr$_{0.5}$NiO$_{4}$ crystals. The x-axis is equal to the inverse of
temperature and the y-axis in presented in logarithmic scale to facilitate
the test of an exponential decay law.\bigskip

Figure 4. Magnetization measurements in the La$_{1.35}$Sr$_{1.65}$Mn$_{2}$O$%
_{7}$ crystal between 1.8 and 10 K with applied magnetic field of 20 Oe (a)
and 50 Oe (b). Measurements were done with the magnetic field parallel
(triangles) and perpendicular (squares) to the {\em c} axes. The results
reveal an anisotropy due to the quasi-bidimensional distribution of the
magnetic ions.\bigskip

Figure 5. Magnetization measurements in the La$_{1.5}$Sr$_{0.5}$NiO$_{4}$
crystal between 1.8 and 10 K with applied magnetic field of 1 T (a) and 3 T
(b). Magnetic field was applied parallel (triangles) and perpendicular
(squares) to the {\em c} axes. The small peak close to 2 K, in the parallel
orientation to the {\em c} axes, might be correlated with a gap in the
electronic structure.

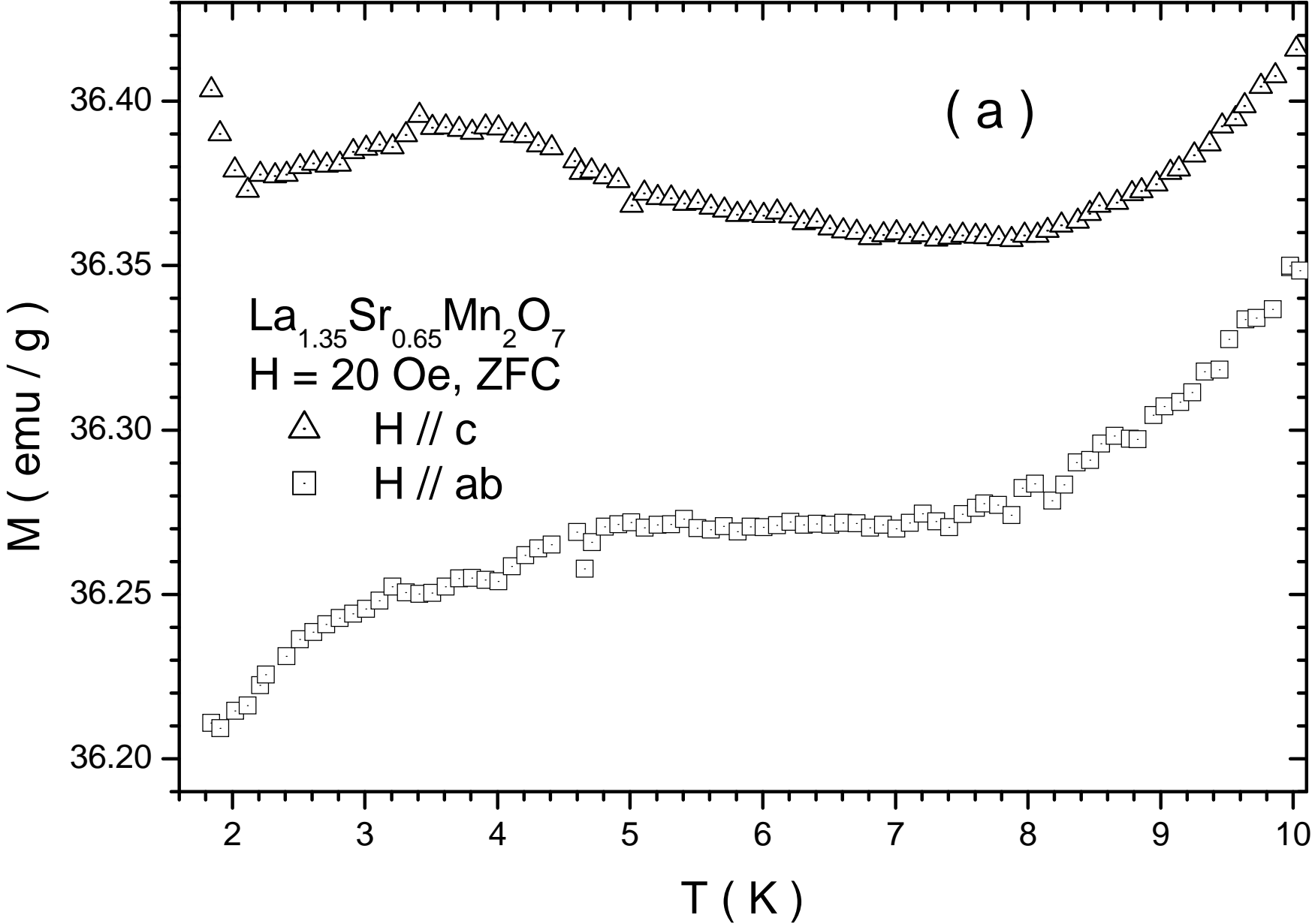
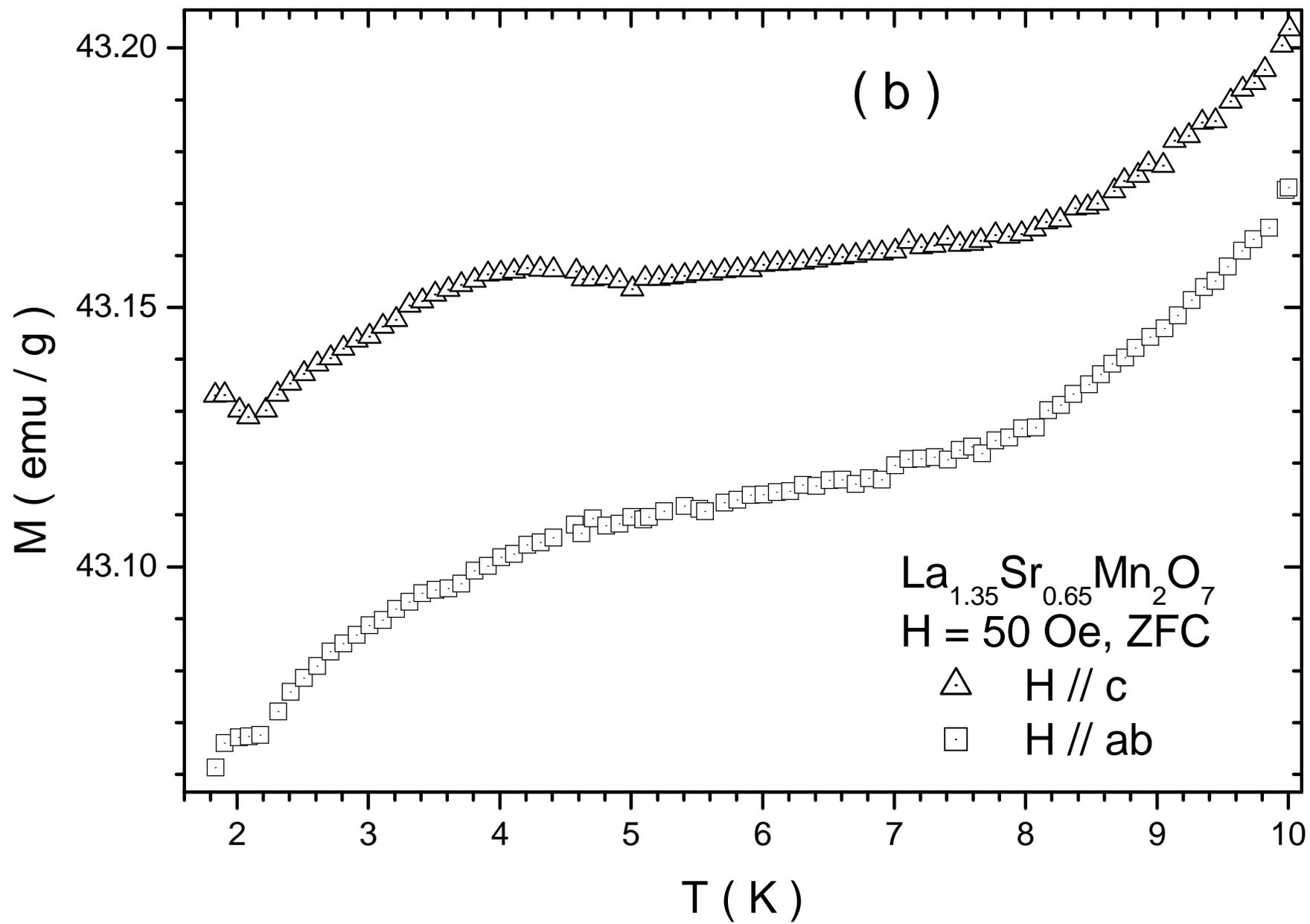
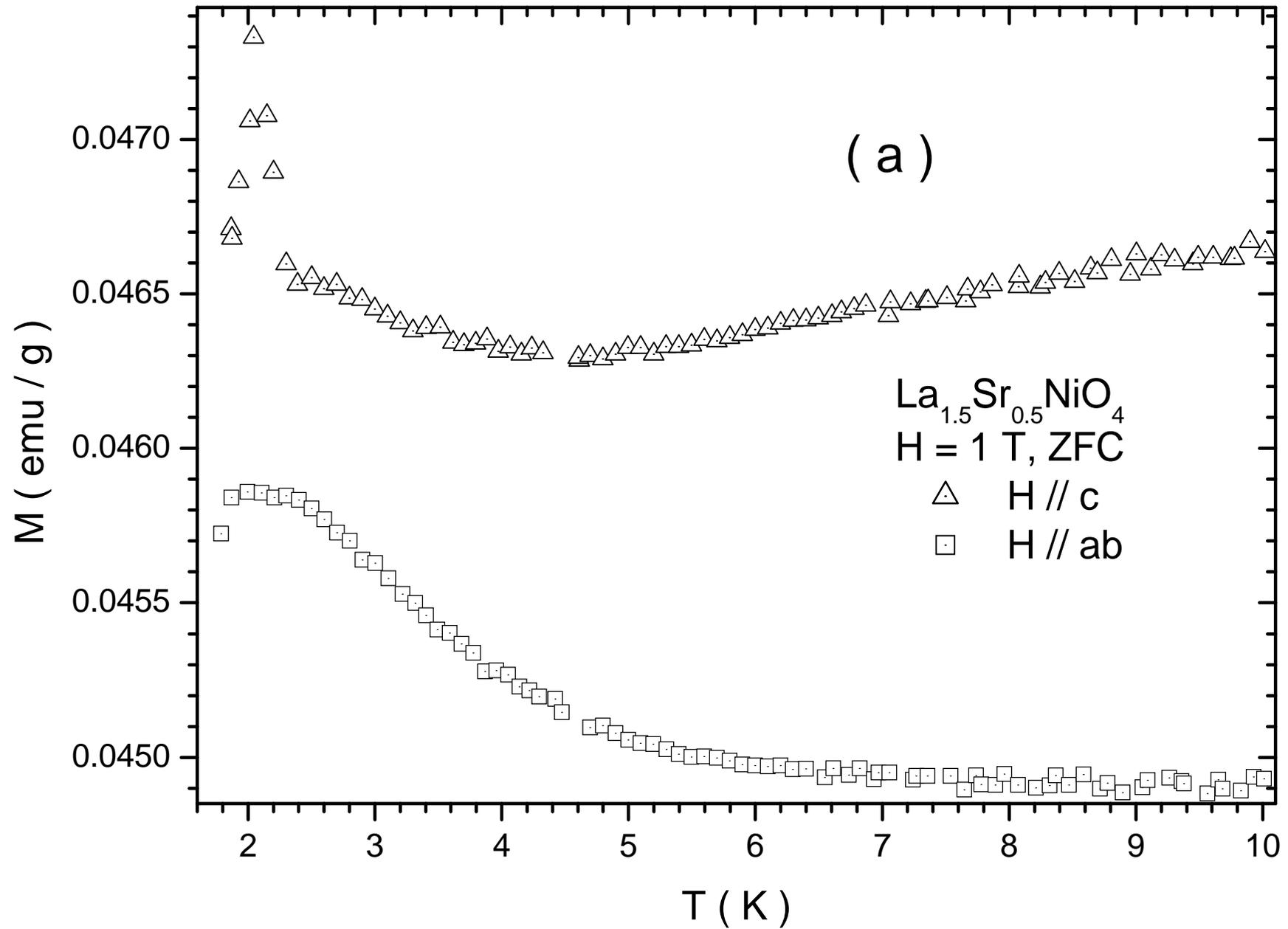
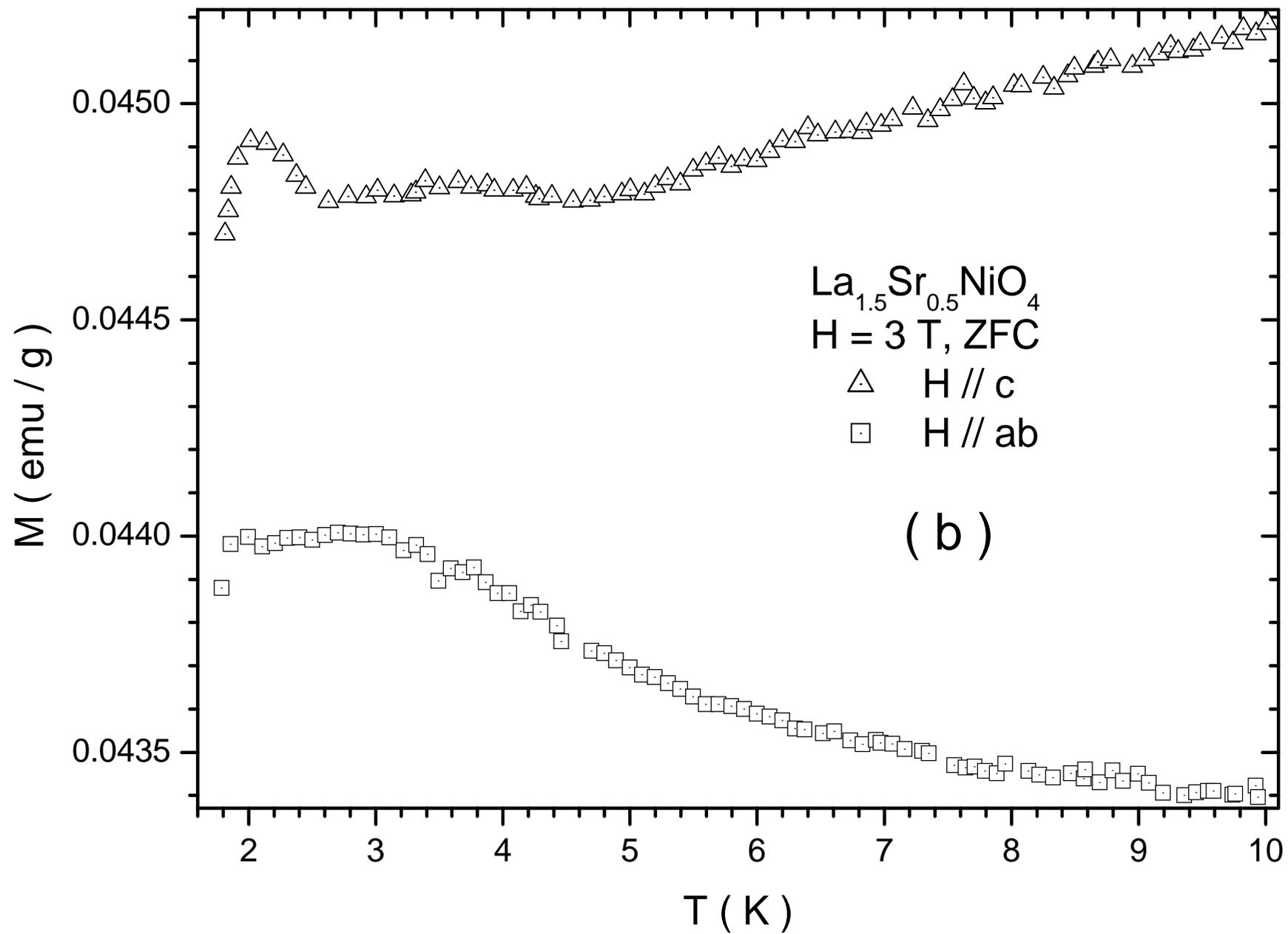
\end{document}